\documentclass[aps,pra,showpacs,twocolumn,amsmath,amssymb,nofootinbib]{revtex4}
\usepackage{graphicx}% Include figure files   
\usepackage{epsfig}
\usepackage{bm}% bold math
\usepackage{amssymb}
\usepackage{amsmath}
\usepackage{latexsym}
\usepackage{longtable}
\usepackage{color}
\newcommand{\vc}[1]{\mbox{\boldmath $#1$}} % vector
\newcommand{\dirate}{{\mathcal D}}   % dispersion rate for a single Bragg reflection
\newcommand{\ind}[1]{_{#1}}    % generalized index 
\newcommand{\indrm}[1]{_{\mathrm {#1}}}    % generalized index instead of the specific
\newcommand{\gd}{g}    % groove density
\newcommand{\beryl}{Al$_2$Be$_3$Si$_6$O$_{18}$}    % groove densitycan be achieved
\newcommand{\gammahp}{\gamma_{\ind{H}}^{\prime}}   % direction cosine for exit waves
\newcommand{\gammah}{\gamma_{\ind{H}}}   % direction cosine
\newcommand{\gammao}{\gamma_{\ind{0}}}   % direction cosine
\newcommand{\Phih}{\Phi^{\prime}}   % angle of K_H to the crystal surface 
\newcommand{\Phio}{\Phi}   % angle of K_0 to the crystal surface
\newcommand{\Phihc}{\Phi^{\prime}_{\ind{0}}}   % angle of K_H to the crystal surface 
\newcommand{\Phioc}{\Phi_{\ind{0}}}   % angle of K_0 to the crystal surface

\newcommand{\phioc}{\phi_{\ind{0}}}

\begin{document}
\title{Diffraction gratings with two-orders-of-magnitude-enhanced dispersion rates for sub-meV-resolution soft x-ray spectroscopy}
\author{Yuri Shvyd'ko} %\email{shvydko@aps.anl.gov}
\affiliation{Advanced Photon Source, Argonne National Laboratory,
  Argonne, Illinois 60439, USA}

\begin{abstract} 
Diffraction gratings with large angular dispersion rates are central
to obtaining high spectral resolution in grating spectrometers
operating over a broad spectral range from infrared to soft-x-ray
domains. The greatest challenge is of course to achieve large
dispersion rates in the short-wavelength x-ray domain.  Here we show
that crystals in non-coplanar asymmetric x-ray Bragg diffraction can
function as high-reflectance soft-x-ray diffraction gratings with
dispersion rates that are at least two orders of magnitude larger than
those that are possible with state-of-the-art man-made gratings. This
opens new opportunities to design and implement soft x-ray resonant
inelastic scattering (RIXS) spectrometers with spectral resolutions
that are up to two orders of magnitude higher than what is currently
possible, to further advance a very dynamic field of RIXS
spectroscopy, and to make it competitive with inelastic neutron
scattering.  We present examples of large-dispersion-rate crystal
diffraction gratings operating near the 930-eV L$_3$ absorption edge
in Cu and of the 2.838-keV L$_3$ edge in Ru.

\end{abstract}

%   
%============================================================
\pacs{41.50.+h,61.05.cp,07.85.Fv, 07.85.Nc}

\maketitle

\section{Introduction}

Diffraction gratings are the most common type of light-dispersing
optical elements. They are essential for optical instruments that
operate in a very broad spectral range from infrared to soft-x-ray
domains.  The shorter the radiation wavelength, the more challenging
it is to manufacture efficient diffraction gratings. Diffraction
gratings are typically not practical at photon energies $E$ above 2
keV, but they are still very useful at lower photon energies in the
soft-x-ray domain, $0.2$~keV~$\lesssim E \lesssim 2$~keV. They are key
optical elements of resonant inelastic x-ray scattering (RIXS)
spectrometers.

RIXS studies in the soft-x-ray domain recently increased tremendously
in importance because researchers discovered that it is possible to
observe and study orbital and magnetic excitations in addition to
charge and lattice collective excitations in condensed matter, notably
in high-$T_{\indrm{c}}$ superconductors \cite{BBB10,LGC11,SWJ12,DDS13}
previously accessible only by neutrons.  This breakthrough became
possible because of rapid improvements in the spectral resolution
$\Delta E$ of RIXS spectrometers at synchrotron radiation facilities
worldwide, initially from $\Delta E \simeq 120$~meV
\cite{GPD06,SSF10}, achieved at SLS in 2006 to recently demonstrated
resolution of $\Delta E \simeq 40$~meV at ESRF \cite{CGP17,BYK18}, and
$\Delta E \simeq 20$~meV at NSLS-II \cite{Bisogni19}.  Various
facilities, such as ALS \cite{WCV14,CSC17}, TLS \cite{Lai14}, DLS
\cite{DLS-RIXS},
MAX IV, 
%See https://www.maxlab.lu.se/veritas for information about the RIXS beam-
%line under construction at MAX IV.
% Project manager: Marcus Agåker (marcus.agaker@maxlab.lu.se)
% Spokesperson: Jan-Erik Rubensson (Jan-Erik.Rubensson@physics.uu.se)
%https://www.maxlab.lu.se/sites/default/files/2014%20Veritas%20poster%20v2-2.pdf
%https://www.maxlab.lu.se/veritas
European XFEL, 
%http://www.xfel.eu/sites/site_xfel-gmbh/content/e63594/e65073/e126274/e136957/Follath_RIXS_eng.pdf
%http://www.xfel.eu/sites/site_xfel-gmbh/content/e63594/e65073/e126274/e135803/Foehlisch_hRIXSCornsortiumTalkWeb_eng.pdf 
% 40,000 resolution
and LCLS-II are now constructing new soft RIXS beamlines and
spectrometers, aiming to achieve higher throughput and higher
resolution. A RIXS spectrometer at NSLS-II has been designed to
achieve a spectral resolution of $\Delta E \simeq 15$~meV, resolving
power $E/\Delta E \simeq 6.6\times 10^4$ \cite{DJB16}.
% SIX: 22-meV demonstrated IXS-2019 @932 eV
% DIAMOND: 35 meV demonstrated IXS-2019 @932 eV

However, to become  competitive with inelastic neutron scattering,
the resolution of the RIXS spectrometers must be improved
by more than an order of magnitude.  Technically, the progress in the
spectral resolution of RIXS became possible due to better focusing of
x-rays, improved spatial resolution of x-ray detectors, and most
importantly by increasing the length of the spectrometer arm from 5~m \cite{GPD06} to
15~m \cite{DJB16}.  However, the potential to further improve 
spectral resolution by increasing spectrometer sizes is approaching
obvious limits.

A more promising approach would be to improve the angular dispersion
rate of the diffraction gratings, which is central to achieving high
spectral resolution of grating monochromators, spectrometers, and
other devices (see Appendix~\ref{spectrometers}). The angular dispersion rate
$\dirate$
measures the angular
variation of the propagation direction of the diffracted photon with
its energy. It is primarily determined by the smallness of the
diffraction grating period, or equivalently by the large value of
groove density $\gd$.  The maximum angular dispersion rates of the
state-of-the-art diffraction gratings is in the range of $\dirate
\simeq 0.2~\mu$rad/meV at $E\simeq 1$~keV, corresponding to groove a density of $\gd \simeq
5~\mu$m$^{-1}$
\cite{GPD06,SSF10,CGP17,BYK18,Bisogni19,WCV14,CSC17,Lai14,DLS-RIXS,DJB16}.
%2500-1800 lines/mm => res power ???100000??? -SIX@NSLS-2
%2700 lines/mm => res power ???? RIXS@DIAMOND

However, nobody can make better gratings than nature. The ``groove''
density of atomic planes in crystals can be as large as one per
crystal lattice period, which is $\gtrsim 1$/nm, i.e., at least two
orders of magnitude larger than what is currently possible with
manmade gratings.  If crystal gratings with atomic groove density
could be efficiently used, this would lead to much higher dispersion
rates and higher-resolution optical devices.

\begin{figure*}[t!]
  \includegraphics[width=0.99\textwidth]{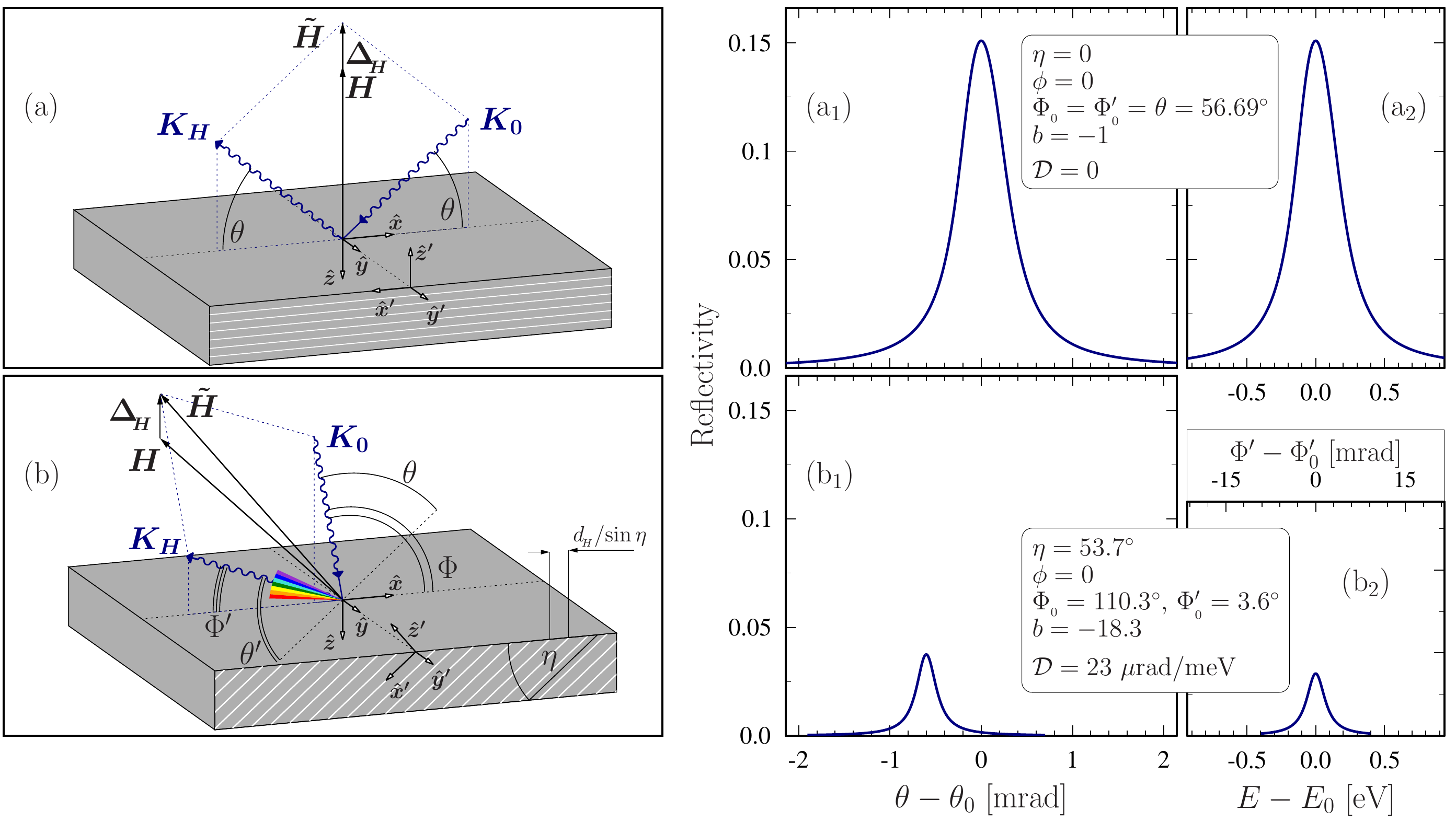}
\caption{Performance of crystals  as soft-x-ray diffraction
  gratings. X-ray reflectivity and dispersion rate $\dirate$  in Bragg
  diffraction from crystals in various scattering geometries (a)-(b)
  as function of the glancing angle of incidence $\theta$ to the Bragg
  diffracting atomic planes (a$_{\ind{1}}$)-(b$_{\ind{1}}$) and of
  x-ray photon energy $E$ (a$_{\ind{2}}$)-(b$_{\ind{2}}$). (a)
  Symmetric geometry with diffracting planes (white parallel lines)
  parallel to the crystal surface. (b) Asymmetric coplanar diffraction
  with diffracting atomic planes at an asymmetry angle $\eta$ to the
  entrance surface, and with the scattering plane
  ($\vc{K}_{\ind{0}},\vc{K}_{\ind{H}}$) parallel to the dispersion
  plane $(\vc{H},\hat{\vc{z}})$. The $\theta$-angle dependences of the
  Bragg reflectivity from the $(10\bar{1}0)$ atomic planes of a beryl
  crystal (\beryl ) are calculated for a photon energy
  $E_{\ind{0}}=930$~eV, and mapped on the angular scale relative to
  $\theta_{\ind{0}}=56.66^{\circ}$, the angular position of the peak
  reflectivity in symmetric Bragg diffraction (a). Each energy $E$
  dependence is calculated at $\theta$ fixed at the  peak
  reflectivity value. The angular dispersion fan [indicated by colors in (b)]
  is in the dispersion plane. Largset $\dirate$ is at largest $\eta$.}
\label{fig001}
\end{figure*}

Here we show that crystals in asymmetric Bragg diffraction, a
scattering geometry with diffracting atomic planes at a nonzero
``asymmetry'' angle $\eta$ to the crystal surface, can function as
high-reflectance ($\simeq 15$\%) broadband ($\simeq 0.5$~eV)
diffraction gratings featuring dispersion rates ($\dirate \simeq
30~\mu$rad/meV at $\lesssim 1$~keV) that are two orders of magnitude
larger than what is possible with manmade gratings.  This may open up new
opportunities to design and implement monochromators and spectrometers
with spectral resolution that is two orders of magnitude higher
($\Delta E \simeq 0.1-1$~meV) in the soft-x-ray regime. This is
especially attractive in view of the rapid recent advances in RIXS
spectroscopy in the soft-x-ray domain calling for further improvements
in spectral resolution.

\section{Angular dispersion in Bragg diffraction}

\begin{figure*}[t!]
  \includegraphics[width=0.99\textwidth]{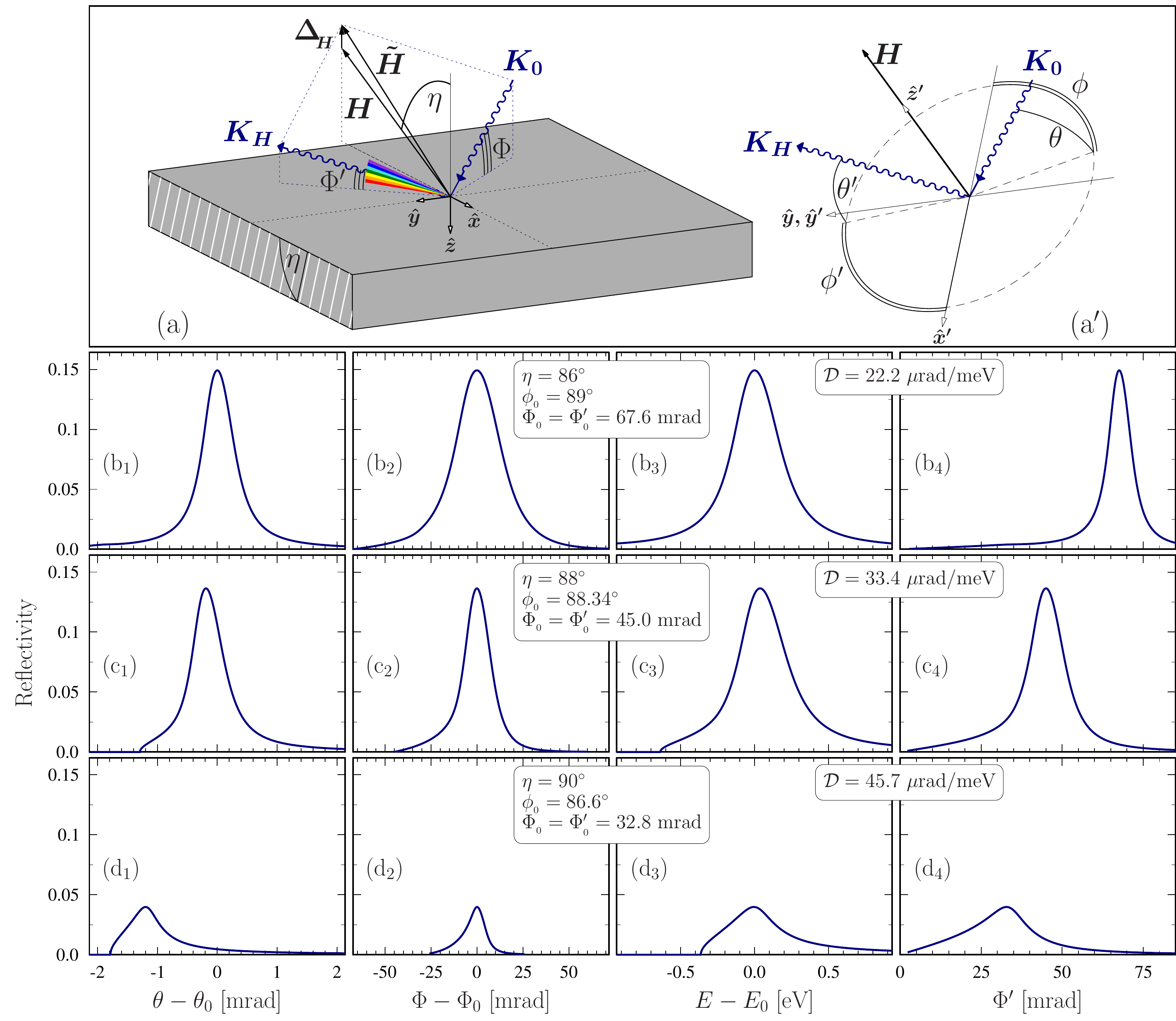}
\caption{Reflectivity and angular dispersion $\dirate$ of soft-x-ray diffraction gratings similar to those in
  Fig.~\ref{fig001}, but in {\em non-coplanar} asymmetric
  scattering geometry as presented in panels (a) and
  (a$^{\prime}$).  In the non-coplanar case, the scattering plane
  ($\vc{K}_{\ind{0}},\vc{K}_{\ind{H}}$) and the dispersion plane
  ($\vc{H},\hat{\vc{z}}$) are no longer parallel. They are
  perpendicular in the case presented here.  The Bragg reflectivity dependences shown in
  rows (b)-(d) are calculated for different asymmetry angles $\eta$.
  In each (b), (c) or (d) case, the azimuthal angle of incidence
  $\phi=\phioc$ (see \cite{Shvydko-SB} for the definition) is chosen
  such that $\Phioc=\Phihc$, where $\Phioc$ and $\Phihc$ are the
  angles between the crystal surface and $\vc{K}_{\ind{0}}$ and
  $\vc{K}_{\ind{H}}$, respectively, as measured at the peak
  reflectivity.  The Bragg reflectivities at fixed photon energy
  $E=E_{\ind{0}}=930$~eV and $\phi=\phioc$ as a function of $\theta$
  and $\Phi$ are shown in (b$_1$)-(d$_1$) and (b$_2$)-(d$_2$),
  respectively.  Bragg reflectivities calculated as a function of $E$!
  with $\theta$ fixed at the peak reflectivity values are presented in
  (b$_3$), (c$_3$), and (d$_3$), while graphs in (b$_4$), (c$_4$), and
  (d$_4$) show the reflectivity mapped on $\Phih$ that changes
  simultaneously with $E$ due to angular dispersion. Larges $\dirate$ is at smallest $\Phioc$ and $\Phihc$.}
\label{fig002}
\end{figure*}

X-ray Bragg diffraction from a crystal alone does not yet guarantee the
grating effect of angular dispersion.  Diffracting atomic planes that
are parallel to the crystal surface [see Fig.~\ref{fig001}(a)] ensure
high x-ray reflectivity from a crystal in Bragg diffraction, but zero
angular dispersion.  The diffracting atomic planes at a nonzero
(asymmetry) angle $\eta$ to the crystal surface [see
  Figs.~\ref{fig001}(b) and \ref{fig002}(a)] ensure both high Bragg
reflectivity and a periodic modulation of the electron density along
the surface, with the latter resulting in the grating effect of the
angular dispersion \cite{Shvydko-SB}. The larger the asymmetry angle,
the shorter the period of the ``diffraction grating'' and the larger
the expected angular dispersion rate.

Here, we study asymmetric Bragg diffraction in coplanar scattering
geometry [shown in Fig.~\ref{fig001}(b)], and in non-coplanar
scattering geometry [presented in Fig.~\ref{fig002}(a)] to reveal the
case featuring both the largest angular dispersion rate and the largest Bragg
reflectivity, along with the largest spectral bandwidth. The
performance of crystals as soft-x-ray diffraction gratings will be
presented on examples of crystals that feature sufficiently large
lattice parameters, such as beryl (\beryl).  

In the coplanar asymmetric scattering geometry, the scattering plane
composed by wave vectors $\vc{K}_{\ind{0}}$ and $\vc{K}_{\ind{H}}$ of
the incident and of the diffracted x-ray photons coincide with the
dispersion plane composed by the diffraction vector $\vc{H}$ and the
internal crystal normal $\hat{\vc{z}}$ [shown in
  Fig.~\ref{fig001}(b)]. In a general case of non-coplanar scattering
geometry they are not parallel. In the particular non-coplanar case
shown in Fig.~\ref{fig002}(a), the scattering and dispersion planes
are perpendicular to each other.

Momentum conservation in Bragg diffraction
\begin{equation}
\vc{K}_{\ind{H}}=\vc{K}_{\ind{0}}+\tilde{\vc{H}}, \hspace{0.5cm} \tilde{\vc{H}}= \vc{H}+\Delta_{\ind{H}}\vc{z}
\label{di0010}
\end{equation}
includes the total momentum transfer $\tilde{\vc{H}}$, that is a sum
of diffraction vector $\vc{H}$ and of an additional momentum transfer
$\Delta_{\ind{H}}$ directed along $\vc{z}$. This is a small but
essential term for the angular dispersion effect, which originates
from refraction at the vacuum-crystal interface \cite{Shvydko-SB}.  It
can be determined from the photon energy $E$ conservation
$|\vc{K}_{\ind{H}}|=|\vc{K}_{\ind{0}}|=K=E/\hbar c$ as
\begin{equation}
\Delta_{\ind{H}}\,=\, K\left( -\gammah\pm\sqrt{\gammah^2-\alpha}\right),
\label{deldel}
\end{equation}
where
\begin{equation}
\gammah=\frac{(\vc{K}_{\ind{0}}+\vc{H})\vc{z}}{{K}_{\ind{0}}}=\gammao -\frac{H}{K}\cos\eta,\hspace{0.5cm} \gammao=\frac{\vc{K}_{\ind{0}}\vc{z}}{{K}_{\ind{0}}}
\label{eq01300}
\end{equation}
are direction cosines of $\vc{K}_{\ind{0}}+\vc{H}$ and of
$\vc{K}_{\ind{0}}$ with respect to $\vc{z}$, respectively; the
dimensionless parameter
\begin{equation}
  \alpha\, =\,
  %  \frac{(\vc{K}_{\ind{0}}+\vc{H})^2-\vc{K}_{\ind{0}}^2}{K^2}=
  \frac{H}{K}\left(\frac{H}{K}-2\sin\theta\right),
%  \frac{2\vc{K}_{\ind{0}}\vc{H}+ \vc{H}^2}{K^2}.
\label{eq32}
\end{equation}
measures deviation from
Bragg's law
$2K\sin\theta=H$; and $\theta$ is a glancing angle
of incidence to the reflecting atomic planes.
The sign in Eq.~\eqref{deldel} is chosen such that
$\Delta_{\ind{H}}=0$ if $\alpha =0$. In the Bragg-case reflection
geometry considered here, in which $\gammah<0$, the sign in
Eq.~\eqref{deldel} is negative (as opposed to the Laue-case
transmission geometry in which $\gammah>0$).

%~/laTeXdocs/papers/q150/noncoplanar-bragg-diffraction.tex}.

The angular dispersion rate $\vc{\dirate}={\mathrm
  d}\vc{u}_{\ind{H}}/{\mathrm d}E$ is calculated using
Eqs.~\eqref{di0010}-\eqref{eq32} as a change with photon energy $E$
(or equivalently with $K$) of the direction of the normalized wave
vector $\vc{u}_{\ind{H}}=\vc{K}_{\ind{H}}/K$ of the diffracted photon
assuming {\em a fixed direction of the incident photon wave vector}
$\vc{K}_{\ind{0}}$:
\begin{gather}
  %\delta\vc{u}_{\ind{H}}\,=\,-\frac{\vc{H}}{K}\,\frac{\delta E}{E} \,+\,  \frac{2\sin^2\theta}{\gammahp}\,\frac{\delta E}{E}, \hat{\vc{z}},
\vc{\dirate}\,=\,-\frac{\vc{H}}{K\,E}\, \,+\,  \frac{2\,\sin^2\theta}{E\,\gammahp}\, \hat{\vc{z}},  
\label{di001}\\
\gammahp\,=\,\pm \sqrt{\gammah^2 -\alpha}\,=\,\frac{\vc{K}_{\ind{H}}\vc{z}}{{K}_{\ind{0}}},
\label{di0014}
\end{gather}
where $\gammahp$ is the cosine of the angle between the vacuum wave
vector $\vc{K}_{\ind{H}}$ and the surface normal $\hat{\vc{z}}$. In
the Bragg-case scattering geometry considered here, $\gammahp< 0$.
Alternatively, $\gammahp$ can be expressed as $\gammahp=-\sin\Phih$
through the sine of an angle $\Phih$ between $\vc{K}_{\ind{H}}$ and
the crystal surface.  Similarly, $\gammao=\sin\Phio$ can be expressed
as the sine of an angle $\Phio$ between $\vc{K}_{\ind{0}}$ and the
crystal surface.  The angles $\Phih$ and $\Phio$ are related to each
other as
\begin{equation}
(\sin\Phih)^2\,=\, \left(\Psi - \sin\Phio \right)^2 -\alpha, \hspace{0.5cm} \Psi=\frac{H}{K}\cos\eta ,
\label{eq01500}
\end{equation}
which follows from Eqs.~\eqref{eq01300} and \eqref{di0014}.

Combining Eqs.~\eqref{di001}-\eqref{di0014} we obtain
\begin{equation}
%  \delta\vc{u}_{\ind{H}}\,=\,\vc{\dirate}\, \delta E,\hspace{0.5cm}
  \vc{\dirate}= -\frac{2\sin\theta}{E} \,\left(\,\frac{\sin\theta}{\sin\Phih}\,\hat{\vc{z}}\,+\, \hat{\vc{z}}^{\prime}\right). 
\label{di00102}
\end{equation}
According to Eq.~\eqref{di00102}, the angular dispersion rate vector
$\vc{\dirate}$ is always in the dispersion plane
$(\hat{\vc{z}},\hat{\vc{z}}^{\prime})$ \cite{Shvydko-SB}. The largest
magnitude $|\vc{\dirate}|$ of the angular dispersion rate  is achieved at
grazing emergence, when the wavevector $\vc{K}_{\ind{H}}$ makes a
small angle, $\Phih\ll 1$, with the crystal surface.

In this extreme case, Eq.~\eqref{di00102} simplifies to
\begin{equation}
\vc{\dirate}\,\simeq\,-\hat{\vc{z}}\,|\dirate| ,\hspace{1cm} |\dirate| \,\simeq \, \frac{1}{E}\frac{2\sin^2\theta}{\sin\Phih}, 
\label{di00200}
\end{equation}
with the dispersion rate vector $\vc{\dirate}$ perpendicular to the
crystal surface\footnote{The angular dispersion rate given by
  Eq.~\eqref{di00200} is equivalent to the angular dispersion rate
  $\dirate \,=\, (g\, hc\,)/(E^2 \sin\Phih)$ of a grating with a
  groove density $g$ (see Appendix~\ref{spectrometers}), which is
  $g=\sin\theta/d_{\ind{H}}$ in our case [see Fig.~\ref{fig001}(b)]
  where the crystal ``grating'' is determined by the interplanar
  distance $d_{\ind{H}}=2\pi/H$ between the diffracting atomic planes
  and incidence angle $\theta$.}. In the angular dispersion fan,
photons with higher energies propagate at larger angles $\Phih$ to the
surface, as indicated by the violet color in Figs.~\ref{fig001}(b) and
\ref{fig002}(a). Using Eq.~\eqref{di00200}, we estimate
$\dirate>20~\mu$rad/meV for $E=930$~eV, assuming $\Phih\lesssim
60$~mrad \footnote{The $\Phih$ value is chosen here to be as small as
  possible, but larger than a critical angle of $\simeq 33$~mrad for
  930~eV photons in beryl.}, the values, which are two orders of
magnitude larger than those possible using manmade diffraction
gratings
\cite{GPD06,SSF10,CGP17,BYK18,Bisogni19,WCV14,CSC17,Lai14,DLS-RIXS,DJB16}.

\section{Soft-x-ray crystal gratings}

As a result, we are coming to a conclusion that very large dispersion rates
$\dirate$ can be achieved in asymmetric Bragg diffraction from
crystals in various scattering geometries, provided $\Phih\ll
1$. However, a critical question is whether the high reflectivity and
spectral bandwidth of the Bragg reflection can be simultaneously
retained.  Let us consider some particular cases, beginning with the
conceptually simplest one of the coplanar scattering geometry.

Figures~\ref{fig001}(a) and \ref{fig001}(b) show schematics of Bragg
diffraction from a crystal in the symmetric $(\eta=0)$ and asymmetric
$(\eta\not=0)$ coplanar scattering geometries,
respectively. Figures~\ref{fig001}(a$_{\ind{1}}$) and
\ref{fig001}(b$_{\ind{1}}$) show the Bragg reflection profiles related
to these scattering geometries as a function of incidence angle
$\theta$ at a fixed photon energy $E=E_{\ind{0}}=930$~eV in the
($10\bar{1}0$) Bragg diffraction from beryl crystal (\beryl ).
Figures~\ref{fig001}(a$_{\ind{2}}$) and \ref{fig001}(b$_{\ind{2}}$)
show the corresponding Bragg reflection profiles as a function of
photon energy $E$, calculated with $\theta$ fixed at the  peak
  reflectivity value.

\begin{figure*}[t!]
  \includegraphics[width=0.99\textwidth]{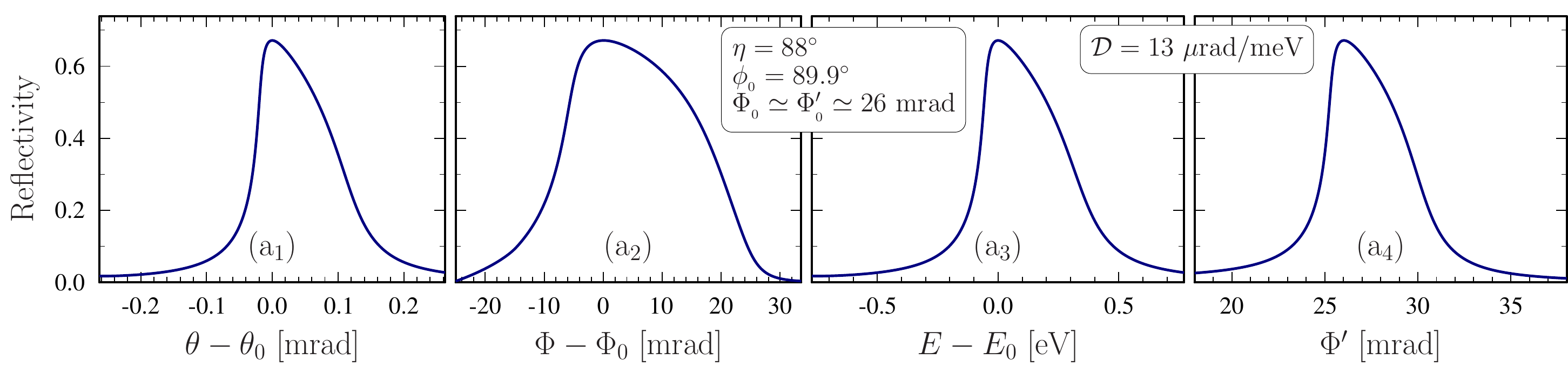}
\caption{Reflectivity and angular dispersion $\dirate$  of x-rays in Bragg diffraction from the (111) atomic
  planes in Si crystal in non-coplanar asymmetric diffraction with
  asymmetry angle $\eta=88^{\circ}$ and
  $\theta_{\ind{0}}=44.16^{\circ}$, similar to those in
  Fig.~\ref{fig002}, but with photon energies in the vicinity of Ru
  $L_{\ind{3}}$ edge ($E_{\ind{0}}$=2.838~keV). }
\label{fig003}
\end{figure*}
  
In the symmetric scattering geometry, the reflection profiles are
relatively broad with an angular acceptance of 0.6~mrad and an energy
bandwidth of 0.37~eV, respectively, featuring about 15\% reflectivity
[see Figs.~\ref{fig001}(a$_{\ind{1}}$)-(a$_{\ind{2}}$)].  Although the
Bragg reflectivity in this case is relatively low compared to the
80\% to 90\% reflectivity typical in the hard x-ray regime (because of high
photoabsorption of the 930~eV photons), it is still larger than the
typical diffraction gratings reflectivity of $\simeq 5$\%. The angular
dispersion rate is, however,  zero in the case of symmetric diffraction.

In contrast, a very large value of the dispersion rate, $\dirate
\simeq 23~\mu$rad/meV, is achieved in strongly asymmetric diffraction
with an asymmetry angle of $\eta=53.7^{\circ}$. In this case the
reflected photons are propagating at varying with photon energy
reflection angle $\Phih$, in particular with $\Phihc=63$~mrad
(3.6$^{\circ}$) at the reflectivity peak; see
Figs.~\ref{fig001}(b$_{\ind{1}}$)-(b$_{\ind{2}}$).  The dispersion
rate is two orders of magnitude larger than that possible with manmade
diffraction gratings at the same photon energy
\cite{GPD06,SSF10,CGP17,BYK18,Bisogni19,WCV14,CSC17,Lai14,DLS-RIXS,DJB16}.
However, this comes at a high price of a factor of three reduced angular
acceptance to 0.24~mrad and of the energy bandwidth to 0.16~eV. The
reflectivity is reduced even more, by a factor of four, to 3.9\%.
%(although this value is close to the reflectivity of the diffraction gratings in the same spectral range).

The drastic reduction of the angular and spectral widths together with
the Bragg reflectivity is due to the large change in the magnitude of
the asymmetry factor, $b=\gammao/\gammah=-\sin\Phio/\sin\Phih$. It
changes from the favorable for the high reflectivity value $b=-1$ in
the symmetric scattering geometry, for which
$\Phio=\Phih=\theta=56.66^{\circ}$, to $b\simeq -18$ in the considered
here asymmetric scattering geometry, for which
$\Phio=\theta+\eta=130.3^{\circ}$ and $\Phih=\theta-\eta=3.6^{\circ}$.

In the following, we show that asymmetric Bragg diffraction can be
also realized featuring $b\simeq -1$ and ensuring not only the large
dispersion rates $\dirate$ as in Fig.~\ref{fig001}(b), but also broad
spectral widths together with high reflectivity, comparable to those
of the symmetric case presented in
Figs.~\ref{fig001}(a$_{\ind{1}}$)-(a$_{\ind{2}}$). This is achieved in
a particular asymmetric non-coplanar scattering case, in which
$\Phio\simeq \Phih$ (``quasi-symmetric'') [see
  Figs.~\ref{fig002}(a)-(a$^{\prime}$)].

Rows x=b,c, and d in
Figs.~\ref{fig002}(x$_{\ind{1}}$)-\ref{fig002}(x$_{\ind{4}}$) show
examples of the angular and spectral Bragg reflection profiles similar
to those presented in Fig.~\ref{fig001}. However, they are calculated
in the non-coplanar ``quasi-symmetric'' scattering geometry with
values of $\Phioc\simeq \Phihc$ decreasing from (a) to (c). The
details of the calculated profiles are explained in the caption of
Fig.~\ref{fig002}.  Importantly, the $\Phio$ and $\Phih$-dependences
presented in Figs.~\ref{fig002}(x$_{\ind{2}}$) and
\ref{fig002}(x$_{\ind{4}}$) show mrad-broad angular acceptances in
$\Phio$ and mrad-broad angular spread in $\Phih$, respectively. The
latter is due to large angular dispersion rate values growing to
$\dirate > 33~\mu$rad/meV with decreasing values of $\Phioc\simeq
\Phihc$.  The Bragg reflectivity is as high and the reflection
profiles are as broad as in the symmetric case; compare
Figs.~\ref{fig002}(x$_{\ind{1}}$) and \ref{fig002}(x$_{\ind{3}}$) with
Figs.~\ref{fig001}(a$_{\ind{1}}$)-\ref{fig001}(a$_{\ind{2}}$). Only
when $\Phioc\simeq \Phihc\simeq 33$~mrad becomes comparable to the
critical angle and the two-beam Bragg diffraction transforms into the
four-beam grazing-incidence diffraction \cite{AM83,AAS84e,BO01} the
reflectivity drops [see
  Figs.~\ref{fig002}(d$_{\ind{1}}$)-\ref{fig002}(d$_{\ind{4}}$)].

Figures~\ref{fig003}(a)-\ref{fig003}(a$_{\ind{4}}$) show another
interesting case of non-coplanar ``quasi-symmetric'' Bragg
diffraction.  Here, the (111) Bragg reflection of 2.838-keV x-rays
from Si crystal features a $\dirate=13~\mu$rad/meV dispersion rate.
The optical element with such a large dispersion rate and a $>
60$\%-reflectivity can be used as an efficient diffraction grating in
a Ru $L_{\ind{3}}$-edge RIXS spectrometer ensuring meV or even sub-meV
resolution.

No doubt, Si-based diffraction gratings will perform well because
nearly flawless synthetic crystals are available. However, Si crystals
are not applicable in the photon spectral range below 2~keV, because
they have a relatively small crystal lattice parameter.  Crystals with
larger lattice parameters are known and have been used in soft-x-ray
Bragg diffraction monochromators
\cite{Johnson83,FSP86,KGG89,XZP98,WDM12}. Among them are beryl
crystals (\beryl), which can diffract photons with energies above
780~eV; KAP crystals diffracting above 600~eV, and others. However,
because their quality is inferior it is of course a question of
whether these crystals can perform appropriately as diffraction
gratings.  In Appendix~\ref{beryl}, we show that natural beryl
crystal quality may be sufficient for them to be used as diffraction
gratings in meV-resolution spectrometers.

\section{Summary}

In summary, we identified an asymmetric Bragg diffraction scattering
geometry that ensures very high angular dispersion rates, along with
high reflectivity and a large spectral reflection range. This is
exemplified in a non-coplanar case, where the scattering and
dispersion planes are perpendicular to each other, and x-rays
propagate are at very small but equal grazing angles of incidence and
reflection to the crystal surface.  The Bragg diffraction dispersion
rates in the soft-x-ray regime can be more than two orders of
magnitude larger than what is possible with manmade diffraction
grating. This creates a path toward the realization of soft-x-ray
spectrometers with resolving power of $10^{7}$ compared to the
currently available $\gtrsim 3\times 10^4$, which will make RIXS
compatible with inelastic neutron scattering.

\section{Acknowledgments}

Tomasz Kolodziej is acknowledged for help with characterization of
beryl crystals by x-ray rocking curve imaging. Xianrong Huang is
acknowledged for determining crystal orientation using Laue method. Work
at Argonne National Laboratory was supported by the U.S. Department of
Energy, Office of Science, Office of Basic Energy Sciences, under
contract DE-AC02- 06CH11357.

%\bibliography{/home/shvydko/laTeXdocs/bibliography/mybib}

%\end{document}

\appendix

\section{Diffraction grating dispersion rate and spectral resolution of soft-x-ray spectrometers}
\label{spectrometers}

The variation of grating groove density $\gd(x)$ in a general case is
represented as a polynomial \cite{WCV14}
\begin{equation}
\gd(x)\,=\,\gd_{\ind{0}}\,+\,\gd_{\ind{1}}\,x\,+\,\gd_{\ind{2}}\,x^2\,+\,\gd_{\ind{3}}\,x^3.
\label{dg0010}
\end{equation}
The plane grating equation is
\begin{equation}
\sin\alpha\,+\,\sin\beta\,=\,m\,\lambda\, \gd_{\ind{0}}\,=\,m\,\gd_{\ind{0}}\frac{hc}{E},
\label{dg0020}
\end{equation}
where $\alpha>0$ and $\beta<0$ are angles of incidence and diffraction, respectively; $m$ is diffraction order; $\lambda$ is wavelength;  $E=hc/\lambda$ is photon energy, and $hc=1.2398$~eV$\cdot\mu$m.

The angular dispersion rate $\dirate$  (variation of $\beta$ with $E$ at fixed $\alpha$) is
\begin{equation}
\dirate\,=\,\frac{{\mathrm d}\beta}{{\mathrm d}E}\,=- \frac{m\,\gd_{\ind{0}}\, hc}{E^2 \cos\beta}.
\label{dg0030}
\end{equation}
With the Au-coated grating parameters $\gd_{\ind{0}}= 5/\mu$m;
$\beta=-87.6^{\circ}$; $m=-1$ provided in \cite{WCV14} the dispersion
rate at $E=930$~eV is $\dirate=0.17~\mu$rad/meV. Gratings with similar
dispersion rates were considered in application to state-of-the-art
major soft-x-ray RIXS spectrometers
\cite{GPD06,SSF10,CGP17,BYK18,Bisogni19,WCV14,CSC17,Lai14,DLS-RIXS,DJB16}. The
angular dispersion rates of the gratings based on Bragg-diffracting
crystals discussed in this paper, are about two orders of magnitude
larger. This can be used to improve the spectral resolution of the
soft-x-ray spectrometers by at least two orders of magnitude.

\begin{figure*}[t!]
  \includegraphics[width=0.999\textwidth]{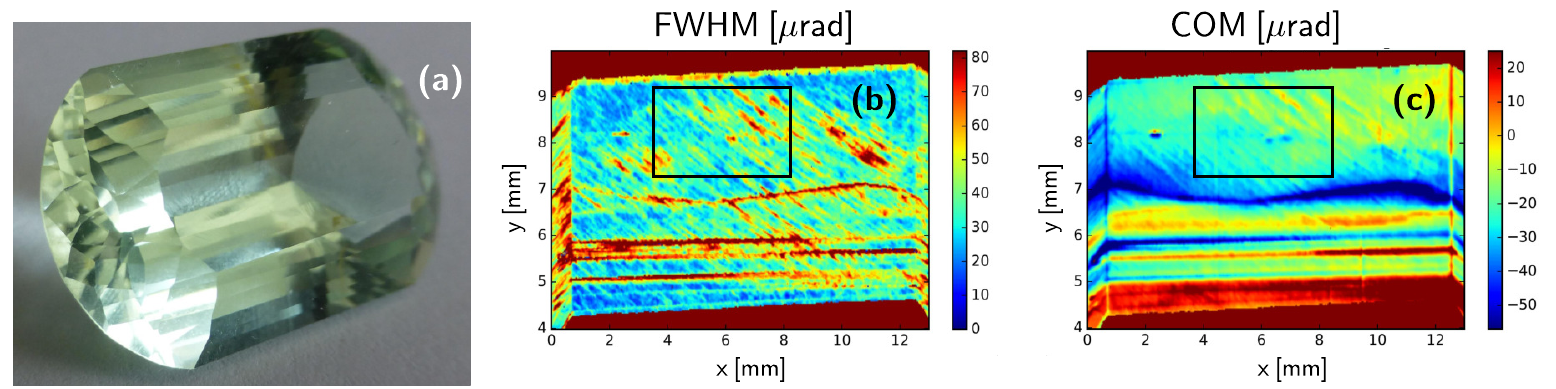}
\caption{A photograph of a natural faceted beryl crystal (a)  characterized by
  x-ray rocking curve imaging (b)-(c).  The (0004) Bragg reflection color maps of the
  angular widths (FWHM) (b), and of the center
  of mass (COM) of the rocking curves (c).  }
\label{fig004}
\end{figure*}

Contributions to the spectral resolution of a grating
spectrometers are listed below, following \cite{GPD06}.

Assuming unlimited detector resolution and a grating with no slope
errors, the major contribution to the spectral resolution of the
spectrometer is due to the secondary source size (on the sample)
$\Delta x_{\ind{1}}$ in the dispersion plane, which is 
\begin{equation}
\Delta E_{\ind{1}}\,=\, \frac{\Delta x_{\ind{1}} \cos\alpha}{r_{\ind{1}}} \frac{E^2 }{m\,\gd_{\ind{0}}\, hc}\,=\, \frac{\Delta x_{\ind{1}}}{r_{\ind{1}}} \, \frac{1}{|\dirate|} \,\frac{\cos\alpha}{\cos\beta}.
\label{dg0040}
\end{equation}
Here, $r_{\ind{1}}$ is a distance from the secondary source to the
focusing grating.  The right-hand side of Eq.~\eqref{dg0040} coincides
with the spectral resolution of the Bragg-crystal-based grating
spectrometer \cite{Shvydko15}, note that ${\cos\alpha}/{\cos\beta}$
has the same meaning as the asymmetry ration $|b|$ in Bragg crystal
optics.

If the monochromatic image size $\Delta x_{\ind{2}}$ on the detector
is smaller than the detector spatial resolution $\Delta
x_{\indrm{D}}$, there is an additional detrimental contribution to the
spectral resolution
\begin{equation}
\Delta E_{\ind{2}}\,=\, \frac{\Delta x_{\indrm{D}} \cos\beta}{r_{\ind{2}}} \frac{E^2 }{m\,\gd_{\ind{0}}\, hc}\,=\, \frac{\Delta x_{\indrm{d}}}{r_{\ind{2}}} \, \frac{1}{|\dirate|}.
\label{dg0050}
\end{equation}
Here, $r_{\ind{2}}$ is a distance from the focusing grating to the
pixel detector.  In this case, the expression also coincides with a
similar expression in \cite{Shvydko15}.

The slope error contribution $\sigma$ (rms) to the spectral resolution
is given by
\begin{equation}
\Delta E_{\ind{3}}\,=\,2.35 \sigma \left(1+\frac{\cos\alpha}{\cos\beta} \right)\, \frac{1}{|\dirate|}.
\label{dg0055}
\end{equation}
The net spectral resolution $\Delta E$ is a sum of all
contributions: $\Delta E=\sqrt{\Delta E_{\ind{1}}^2+\Delta
  E_{\ind{3}}^2+\Delta E_{\ind{3}}^2}$.

It is remarkable that
\begin{equation}
\Delta E_{\ind{1}}\propto \frac{1}{\dirate},~\Delta E_{\ind{2}}\propto \frac{1}{\dirate},~\Delta E_{\ind{3}}\propto \frac{1}{\dirate},~{\mathrm {and}~~} \Delta E\propto \frac{1}{\dirate},
\label{dg0060}
\end{equation}
in other words, a large value of the angular dispersion rate $\dirate$
is favorable for reducing detrimental contributions of each component
and achieving the highest spectral resolution of the
spectrometers.

\section{Crystal quality of natural beryl: x-ray rocking-curve-imaging studies}
\label{beryl}

Feasibility of the Bragg-diffraction soft-x-ray crystal dispersing
elements (diffraction gratings) relies on the availability of crystals
with a large crystal lattice parameter and good crystal
quality. High-quality crystals such as silicon cannnot diffract x-rays
with photon energies below 2~keV.  Crystals with larger lattice
parameters are known and have been used in soft-x-ray Bragg
diffraction monochromators
\cite{Johnson83,FSP86,KGG89,XZP98,WDM12}. Among them are beryl
crystals (\beryl), which can diffract photons with energies above
780~eV; KAP crystals diffracting x-rays above 600~eV, and
others. However, because of their inferior quality it is a question
whether they can perform appropriately as diffraction gratings.

A few natural beryl crystals were obtained from D Sarros Gems Limited
(Geneva, IL, USA) to evaluate their Bragg-diffraction performance.
Samples were selected in the (0001) orientation. Figure~\ref{fig004}(a)
shows a photograph of one of the crystals, with lateral dimensions of
$\simeq 15\times 25$~mm$^2$.

The selected beryl crystals were characterized by sequential x-ray
Bragg diffraction topography \cite{LBH99}, also known as rocking curve
imaging (RCI) with 8-keV x-rays.  This technique measures Bragg
reflection images of a crystal with a pixel x-ray detector
sequentially at different incidence angles to the Bragg reflecting
atomic planes.  The angular dependences of Bragg reflectivity (rocking
curves) measured with each detector pixel are used to calculate Bragg
reflection maps: first, a map of the angular widths [full width at
  half maximum (FWHM)] of the rocking curves is calculated, shown as a
color map in Fig.~\ref{fig004}(b), and, second, a color map of the
center of mass (COM) of the rocking curves shown in
Fig.~\ref{fig004}(c).  The microscopic defect structure can be derived
from the Bragg reflection FWHM maps. The mesoscopic and macroscopic
crystal strain and Bragg planes slope errors can be best evaluated
from the COM maps.

We used a sequential x-ray diffraction topography setup at X-ray
Optics Testing 1-BM Beamline at Argonne's Advanced Photon Source
\cite{SST16}. The setup enables rocking curve mapping with a
submicroradian angular and 13-$\mu$m spatial resolution, limited by
the detector pixel size. The setup employed a close-to-nondispersive
double-crystal Si(220)-Beryl(0004) arrangement with the first
asymmetrically cut high-quality silicon conditioning crystal (Bragg's
angle $\theta_{\ind{220}}=23.8^{\circ}$), and the second beryl crystal
under investigation (Bragg's angle
$\theta_{\ind{0004}}=19.7^{\circ}$).  The expected in theory value of
the 0004 Bragg reflection width in beryl is
$\Delta\theta_{\ind{0004}}=12~\mu$rad (FWHM).  Scales on the $x$- and
$y$-axis in Figs.~\ref{fig004}(b)-(c) correspond to the detector
coordinates. The diffraction plane goes through the $y$-axis.  The
crystal therefore appears to be contracted by a factor of
$\sin\theta_{\ind{0004}}=0.3$ in the $y-$direction.  The maps were
calculated using a dedicated code \cite{Stoupin15}.

The reflection maps show that crystal quality varies
substantially. They also reveal a layered crystal structure, which
indicates that the crystal grew in the earth's crust under varying
conditions over time.  The inclined parallel lines result from crystal
polishing.

In the region indicated by the black rectangle, which is about
4($x$)$\times$6($y$)~mm$^2$ (sufficient in size for our application)
the quality is relatively homogeneous. The averaged over the selected
area reflection width is $\Delta \theta= 23~\mu$rad (FWHM) and the
averaged peak variation is $\sigma_{\indrm{COM}}=4~\mu$rad.  These
results clearly show that the crystals are not perfect. However, these
results are promising, because the measured values of $\Delta \theta$
and $\sigma_{\indrm{COM}}$ have to be compared with the reflection
width and dispersion rates expected in Bragg diffraction of 930-eV
x-rays, which are 0.6~mrad and $22~\mu$rad/meV, respectively. Because
the expected values are larger than the measured ones, the results of
these studies show that the quality of natural beryl crystals may be
sufficient for use as soft-x-ray crystal diffraction
gratings.

We note also that radiation damage of beryl was reported previously
when it was used in an x-ray monochromator under high-heat-load conditions
\cite{FSP86}.  This should not be a problem if beryl crystals are used
as analyzers or diffraction gratings \cite{WDM12}. The grazing
incidence geometry foreseen for the beryl's application as a
diffraction grating should be favorable to mitigate the radiation
load.  In addition, use of cryogenic cooling may improve thermal
conductivity and reduce radiation damage on the monochromator crystal.

\end{document}